\documentclass[twocolumn,twoside,slac_two]{revtex4}
\usepackage{graphicx}
\usepackage{fancyhdr}
\begin{document}

\title{Gravitational Radiation from Coalescing Supermassive Black Hole Binaries in a Hierarchical Galaxy Formation Model}

%

\author{Motohiro Enoki \footnote{enoki.motohiro@nao.ac.jp} ${}^1$, Kaiki T. Inoue${}^2$, Masahiro Nagashima${}^3$ and Naoshi Sugiyama${}^{1}$
}

\affiliation{${}^1$National Astronomical Observatory of Japan, Tokyo 181-8588, Japan}

\affiliation{${}^2$Department of Natural Science and 
Engineering, 
Kinki University, Higashi-Osaka, Osaka, 577-8502, Japan}

\affiliation{${}^3$Department of Physics, 
Kyoto University, Sakyo-ku, Kyoto, 606-8502, Japan}

\begin{abstract}
We investigate the expected gravitational wave emission 
from coalescing supermassive
black hole (SMBH) 
binaries resulting from mergers of their host galaxies.
We employ a semi-analytic 
model of galaxy and quasar formation  based on the
hierarchical clustering scenario to estimate the
amplitude of the expected stochastic 
gravitational wave background owing to inspiraling SMBH binaries 
and bursts rates owing to the SMBH binary coalescence events.
We find that the characteristic
strain amplitude of the background radiation is  
$h_c(f) \sim 10^{-16} (f/1 \mu {\rm Hz})^{-2/3}$ for $f \lesssim 1 \mu {\rm Hz}$.
The main contribution to the total strain amplitude of the background
 radiation comes from SMBH coalescence events at $0<z<1$.
We also find that a future 
space-based gravitational wave interferometer such as the planned 
\textit{Laser Interferometer Space Antenna} ({\sl LISA}) might
detect intense gravitational wave bursts associated with coalescence of SMBH
binaries with total mass $M_{\rm tot} < 10^7 M_{\odot}$ at $z \gtrsim 2$
 at a rate $ \sim 1.0 \ {\rm yr}^{-1}$. 
Our model predicts that burst signals 
with a larger amplitude $h_{\rm burst} \sim 10^{-15}$
correspond to coalescence events of 
massive SMBH binary with total mass $M_{\rm tot} \sim 10^8
M_{\odot}$ at low redshift  $ z \lesssim 1$ 
at a rate $ \sim 0.1 \ {\rm yr}^{-1}$
whereas those with a smaller amplitude $h_{\rm
 burst} \sim 10^{-17}$ correspond to coalescence events of 
less massive SMBH binary with total mass
$M_{\rm tot} \sim 10^6 M_{\odot}$ 
at high redshift $ z \gtrsim 3$.

\end{abstract}

\maketitle

\thispagestyle{fancy}


\section{Introduction}\label{sec:intro}
In recent years, there has been  increasing observational evidence that many nearby galaxies have central supermassive black holes (SMBHs)
in the mass range of $10^6-10^9 M_{\odot}$,
 and that their physical properties
correlate with those of 
spheroids of
their host galaxies. 
This suggests that the
formation of SMBHs physically links to the formation of spheroids that
harbor the SMBHs. Therefore,
in order to study the
formation and evolution of SMBHs, it is necessary to construct a model
that includes galaxy formation processes.

In the standard hierarchical structure formation scenario in a cold dark matter (CDM) universe, dark-matter halos ({\it dark halos}) cluster gravitationally
and merge together. In each of merged dark halos, a galaxy 
is formed as a result of  
radiative gas cooling, star formation, and  
supernova feedback. 
Several galaxies in a common dark halo sometimes merge together and a
more massive galaxy is assembled. When galaxies merge, SMBHs in the
centers of the galaxies sink toward the center of the new merged galaxy and 
form a SMBH binary subsequently. 
If the binary loses enough energy and angular 
momentum, it will evolve to the gravitational wave emitting 
regime and begin inspiraling, eventually coalesces with a 
gravitational wave burst. 

An ensemble of 
gravitational waves from a number of inspiraling 
SMBH binaries at different redshift can be observed
as a stochastic background at frequencies
$\sim 1{\rm n} - 1 \mu {\rm Hz}$, which can be detected by 
pulsar timing measurements.
Future 
space interferometers such as the \textit{Laser Interferometer Space 
Antenna} ({\sl LISA}) might detect quasi-monochromatic 
wave sources associated with inspiraling SMBH binaries 
and gravitational wave bursts associated
with SMBH binary coalescence \citep[e.g.][]{Haehnelt94}. 

To date, a number of attempts have been
made to calculate the SMBH coalescence rate.
Some authors use phenomenological models of
galaxy mergers based on number counts of quasars 
and spheroids \citep[e.g.][]{Thorne76,Jaffe03}. Others use 
merger rates of dark halos (not galaxies) \citep[e.g.][]{Haehnelt94}. 
However, none of these models include baryonic gas evolution 
and galaxy formation processes.
Because SMBH formation process is relevant to spheroids of host
galaxies rather than to dark halos, we need to evaluate 
how the baryonic gas processes such as star formation and supernova
feedback affect the SMBH formation process.

In this study, we estimate the SMBH coalescence rate 
using a new semi-analytic (SA) model \cite{Enoki03}
(an extended model \cite{Nagashima01b})
in which the SMBH formation is incorporated into the galaxy formation. 
Then, we calculate the spectrum of gravitational wave
background from inspiraling SMBH binaries, based on the formulation given by
Jaffe \& Backer \cite{Jaffe03} and we compare our result with 
that from a pulsar timing measurement.
We also estimate the event rate of gravitational wave bursts from
SMBH coalescence events that might be detected by future planned space laser 
interferometers, based on an argument in \cite{Thorne76}. 
The details of formulation, results, discussion and references are given in \cite{Enoki04}.

\section{Galaxy Merger / Black Hole Coalescence Rate}\label{sec:merger}
 Here we briefly describe our SA model for galaxy formation
 and SMBH growth. The details are shown in
 \cite{Nagashima01b} and \cite{Enoki03}. 
 
\subsection{The model of galaxy formation}\label{subsec:galmodel}
First, we construct Monte Carlo realizations of merging histories of
 dark halos from the present to higher redshifts. 
Merging histories of dark halos depend on the 
 cosmological model. The adopted cosmological model is a low-density,
 spatially flat cold dark matter ($\Lambda$CDM) universe with
the present density parameter, $\Omega_{\rm m}=0.3$, the cosmological
constant, $\Omega_{\Lambda}=0.7$, the Hubble constant $h=0.7$ ($h \equiv
 H_0/100 \; {\rm km \
 s^{-1}\ {Mpc^{-1}}}$) and the present rms density fluctuation in spheres
 of $8 h^{-1} {\rm Mpc}$ radius, $\sigma_8=0.9$.
The highest redshift in each merging path which depends on
 the present dark halo mass, is about $z \sim 20-30$. 

Next, in each merging path of dark halos, we calculate the
evolution of the baryonic component from higher redshifts to the
present using simple analytic models for gas cooling,
star formation, supernova feedback, galaxy merging and other processes. 
When a dark halo
 collapses, the gas in the halo is shock-heated to the virial
 temperature of the halo (the {\it hot gas}). At the same time, the gas
 in dense regions of the halo cools owing to efficient 
radiative cooling and sinks to the center of
 the halo and settle into a rotationally supported disk until the
 subsequent collapse of the dark halo.  We call this cooled
gas the {\it cold gas}. 
Stars are formed from the cold gas.
With star formation, supernovae occur
and heat up the surrounding cold gas to the hot gas phase (supernova
feedback). 

When several progenitor halos have merged, a newly formed larger dark
halo contains at least two or more galaxies which had originally
resided in the individual progenitor halos.   
We identify the central galaxy in the new common halo with the central galaxy
contained in the most massive of the progenitor halos.  
Other galaxies are regarded as {\it satellite galaxies}.
After each merging of halo, these satellites merge by either dynamical
friction or random collision. 
Satellite galaxies merge with the central
galaxy in the dynamical friction timescale \cite{BT87}. 
Satellite galaxies sometimes merge with each other in 
the timescale of random collisions. 
Under the condition that the satellite galaxies gravitationally bound 
and merge during encounters, this timescale is derived by Makino \& Hut \cite{Makino97}. 
If the mass ratio, $f=m_1/m_2$, is larger than a certain
critical value of $f_{\rm bulge}$, we assume that a starburst occurs, and
that all of the cold gas turns into stars and hot gas, which fills the
dark halo, 
and all of the stars populate the bulge of a new galaxy.
On the other hand, if $f<f_{\rm bulge}$, no starburst occurs and a smaller 
galaxy is simply absorbed into the disk of a larger galaxy. 

Model parameters are determined  by a comparison with observations of
galaxies in the local Universe, such as 
luminosity functions and the cold gas mass fraction in spiral galaxies.
Our SA model can reproduce galaxy number counts and
photometric redshift distribution of galaxies in the Hubble Deep Field \cite{Nagashima01b}.

\subsection{The growth of SMBHs}\label{subsec:smbhmodel}
\begin{figure*}
\includegraphics[width=95mm]{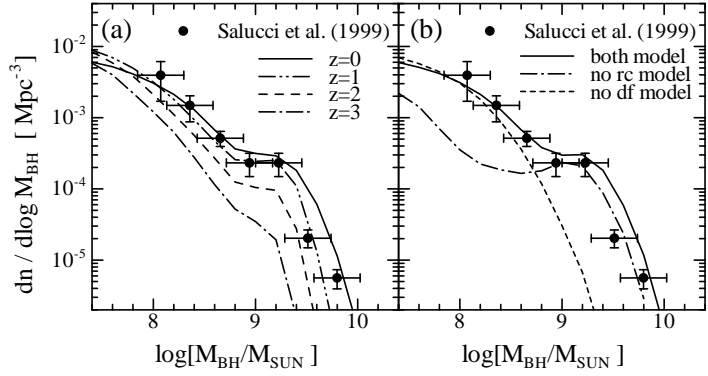}
\caption{(a) Black hole mass function of the model 
 at a series of redshifts. The symbols with
 errorbars are the
 observed black hole mass function at $z=0$ \cite{Salucci99}. 
(b) Black hole mass function at $z=0$ for different galaxy merger models,
 which are both model, no random collision model (no rc model) and no
 dynamical friction model (no df model). Both model
 includes dynamical friction and random collision as the galaxy merging
 mechanism. \label{fig1}}
\end{figure*}

In our model, it is assumed that SMBHs grow by coalescence when their 
host galaxies merge and are fueled by accreted
 cold  gas during major mergers of galaxies. 
 When the host galaxies merge, pre-existing SMBHs
sink to the center of the new merged galaxy due to dynamical friction
(or other mechanisms such as gas dynamics), evolve to
the gravitational wave emission regime and eventually coalesce. 
Although the timescale for this process is unknown, 
for
the sake of simplicity  we assume that SMBHs instantaneously evolve to
the gravitational wave emission regime and coalesce. 
Gas-dynamical simulations have demonstrated that the major merger of
 galaxies can
drive substantial gaseous inflows and trigger starburst activity.
Thus, it is reasonable to assume that 
during a major merger of galaxies,
a certain fraction of the cold gas
that is proportional to the total mass of stars 
newly formed at starburst accretes onto the newly formed SMBH
 and this accretion process leads to a quasar activity. 
Under this assumption, the mass of cold gas accreted on a SMBH is given by  
\begin{eqnarray}  
 M_{\rm acc} &=& f_{\rm BH} \Delta M_{*, \rm burst}, \label{eq:bhaccret}
\end{eqnarray} 
where $f_{\rm BH}$ is a constant and $\Delta M_{*, \rm burst} $ is the
total mass of stars formed at starburst. The free parameter of $f_{\rm BH}$ is fixed 
by matching  the observed relation between a spheroid mass and
a black hole mass  $M_{\rm BH} /M_{\rm spheroid} = 0.001 - 0.006$ (e.g. \cite{Magorrian98}); we find that the favorable value of
$f_{\rm BH}$ is nearly $0.03$. 

Figure \ref{fig1} (a) shows the black hole mass functions in our
model at a series of various redshifts. 
In this figure, we superpose the observed black hole mass
function at $z=0$ \cite{Salucci99}. 
The predicted mass function is quite consistent with the
observation. 
Our galaxy formation model
includes dynamical friction and random collision as galaxy merging
 mechanisms. For comparison, in figure \ref{fig1} (b), we also plot
 the black hole mass functions at $z=0$ of other
two models: no random collision model (no rc model) and no
 dynamical friction model (no df model). In the no rc model and the 
no df model,
mergers owing to random collision and dynamical friction, respectively, are
switched off.
 This figure shows that the mass function for low mass 
black holes are determined by random
collisions between satellite galaxies and that for high
mass black holes are influenced by dynamical friction.
The shape of black hole mass function depends on detailed gas processes.
The important contribution of mass increment of SMBHs in central galaxies is
 the cold gas, which is accreted to only central galaxies. 
SNe feedback removes this cold
gas more efficiently in smaller galaxies with $V_{\rm circ} < 280 {\rm km~s}^{-1}$ ($M_{\rm gal} < 10^{12} M_{\odot}$). Thus, the growth of the SMBHs in small
galaxies suffers from SNe feedback. In the no rc model, SMBHs
mainly exist central galaxies. Therefore, the shape of the
black hole mass function in the no rc model has a bump at high mass end
($M_{\rm BH} \sim 10^{9} M_{\odot}$) which corresponds to SMBHs in the
central galaxies ($M_{\rm gal} \sim 10^{12} M_{\odot}$). 
On the other hand, in the no df model, high mass SMBHs cannot be produced since
galaxies cannot merge with the massive central galaxy.

Using the SA model incorporated with this SMBH growth model, we estimate
the comoving number density, $n_c(M_1,M_2,z) dM_1 dM_2 dz$, 
of the coalescing SMBH binaries with mass $M_1 \sim M_1 + dM_1$ and $M_2
\sim M_2 +dM_2$ at $z$ in a given redshift interval $dz$. 

\section{Gravitational Radiation}\label{sec:GWR}

\subsection{Background radiation from SMBH binaries}\label{subsec:gwbg}
In order to calculate the spectrum of gravitational wave background radiation
from SMBH binaries, 
we improve the  formulation of Jaffe \& Backer \cite{Jaffe03}. The details are described
in \cite{Enoki04}.
The spectrum of the gravitational wave background
radiation which  we finally obtain is
\begin{eqnarray}
  h^2_c(f)
& = & \int dz \; dM_1 \; dM_2 \; \frac{4 \pi c^3}{3}
 \left(\frac{G M_{\rm chirp}}{c^3} \right)^{5/3} (\pi f )^{-4/3} \nonumber \\
& & \times  (1+z)^{-1/3}
 n_c(M_1,M_2,z) \; \theta (f_{\rm max} - f), 
\label{eq:spectrum}
\end{eqnarray}
where ${M_{\rm chirp}}=[M_1 M_2 (M_1 + M_2)^{-1/3}]^{3/5}$ is the chirp
mass of the system and $c$ is the speed of light, $f$ is the observed
frequency of the gravitational wave from the binary in a circular orbit,
the $f_{\rm max}$ is maximum  frequency 
and $\theta(x)$ is the step function. 
As a SMBH binary evolves with time, the 
frequency becomes higher. We assume that 
the binary orbit is quasi-stationary  until
the radius equals to $3 R_{\rm S}$, where $R_{\rm S}$ is the Schwarzschild
radius : the radius of the innermost stable circular orbit (ISCO) for
a particle and a non-rotating black hole. 
Then $f_{\rm max}$ is 
\begin{eqnarray}
f_{\rm max}(M_1,M_2,z) &=& \frac{c^3}{6^{3/2} \pi G M_1 (1+z)} \left(1+\frac{M_2}{M_1} \right)^{1/2} \nonumber \\
&=& 4.4 \times 10^{-5} (1+z)^{-1} \left(\frac{M_1}{10^8 M_{\odot}}\right)^{-1} \nonumber \\
& & \times  \left(
1+\frac{M_2}{M_1} \right)^{1/2} {\rm Hz}, \label{eq:fmax}
\end{eqnarray}
where $M_1$ and $M_2$ are SMBH masses ($M_1 > M_2$).

\begin{figure}
\includegraphics[width=85mm]{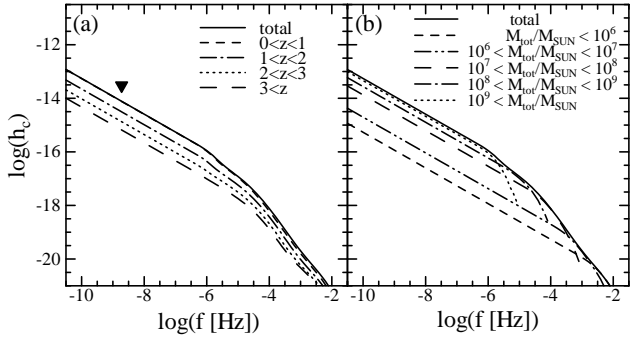}
\caption{Spectrum of gravitational wave background radiation, $h_{\rm c} (f)$, from SMBH binaries in different redshift intervals (a) and in different total mass intervals (b). 
The filled reverse triangle in (a)
 shows the current limit from pulsar timing measurements \cite{Lommen02}.
 \label{fig2}}
\end{figure}

As shown in figure \ref{fig2}, 
the spectrum changes its slope at $f \sim 1 \mu {\rm Hz}$
owing to lack of power associated with the upper limit frequency,
$f_{\rm max}$. This feature is consistent with the results of Sesana et
al. \cite{Sesana04}.
The predicted  strain spectrum is $h_{\rm c}(f) \sim 10^{-16} (f/1 \mu
{\rm Hz})^{-2/3}$ for $f \lesssim 1 \mu {\rm Hz}$, just
below the current limit from the pulsar timing measurements \cite{Lommen02}. 
In our model, we assume that SMBHs coalesce simultaneously when their host
galaxies merge. Therefore, the efficiency of SMBH coalescence is maximum
and the predicted amplitude of gravitational wave spectrum should be
interpreted as the upper limit. 

In figure \ref{fig2} (a), we plot the spectra from binaries in
different redshift intervals. This figure shows that the total spectrum
of background radiation comes from coalescing SMBH binaries 
at low redshift, $0 \le z<1$. 
Although the coalescence rate at low redshift is lower than  at high redshift, 
the main contribution to the  background radiation is 
the coalescing SMBH binaries at low redshift, $0 \le z<1$. This is 
 because the distance from an observer to the SMBH binaries at low redshift is
shorter and the mass of SMBHs at low redshift is higher.
In figure \ref{fig2}(b), we also plot the spectra from binaries in different
 total mass intervals ($M_{\rm tot} = M_1+M_2$). This figure shows that for $f \gtrsim
 10^{-4} {\rm Hz}$ the total
 spectrum of background radiation comes from coalescing SMBH binaries
 with total mass $M_{\rm tot} \lesssim 10^8 M_{\odot}$. 

\begin{figure}
\includegraphics[width=85mm]{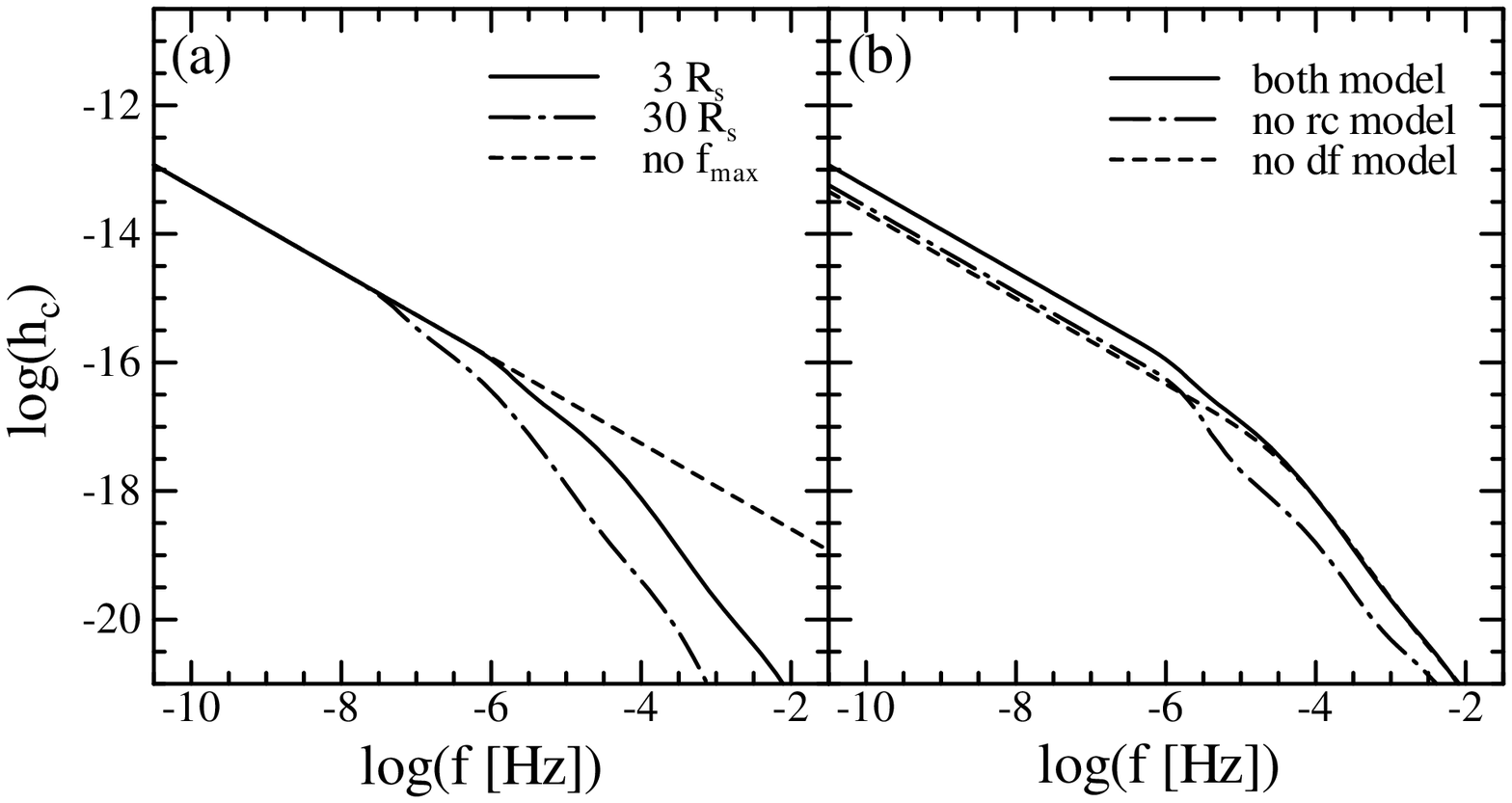}
\caption{Spectrum of gravitational wave background radiation, $h_{\rm c} (f)$,
in different $f_{\rm max}$ (a) and in different galaxy merger models (b). 
\label{fig3}}
\end{figure}

When SMBHs are spinning and/or when their
masses are comparable, the definition of ISCO becomes vague and
our assumption that the cutoff frequency $f_{\rm max}$ corresponds to 
$3 R_{\rm S}$ may not be correct. 
To see the effects of the cutoff frequency, we plot the spectra for
different values of $f_{\rm
max}$, corresponding to $3  R_{\rm S}$, $30 R_{\rm S}$, and no frequency
cut off, respectively (fig. \ref{fig3} (a)). Lowering $f_{\rm
max}$ causes a suppression in the stochastic 
background at high frequencies $f\lesssim 10^{-7}$Hz 
since a large portion of high frequency modes are cut off. 

Our galaxy formation model incorporates dynamical friction and random
collision as galaxy merging mechanisms. In order to examine the effect 
of these two galaxy merger mechanisms on the spectrum of gravitational 
wave background radiation, in figure \ref{fig3} (b), we also plot the 
spectrum of background radiation of other two models: no rc model and no
df model. 
 The no df model can not produce higher mass SMBH (see
 fig. \ref{fig1}(b)). Consequently, the spectrum in no df model bends
 at larger frequency since the number of SMBHs with smaller  $f_{\rm max}$
  decreases. 
Furthermore, in the no df model,
 the amplitude of gravitational waves from each binary becomes smaller
. 
However, the coalescence rate in no df model is 
 higher than the rate in no rc model. 
As a result, the amplitude of the spectrum in no df model is
 roughly equal to the amplitude in no rc model.

\subsection{Gravitational wave bursts from SMBH coalescence}\label{subsec:burst}
\begin{figure*}
\includegraphics[width=95mm]{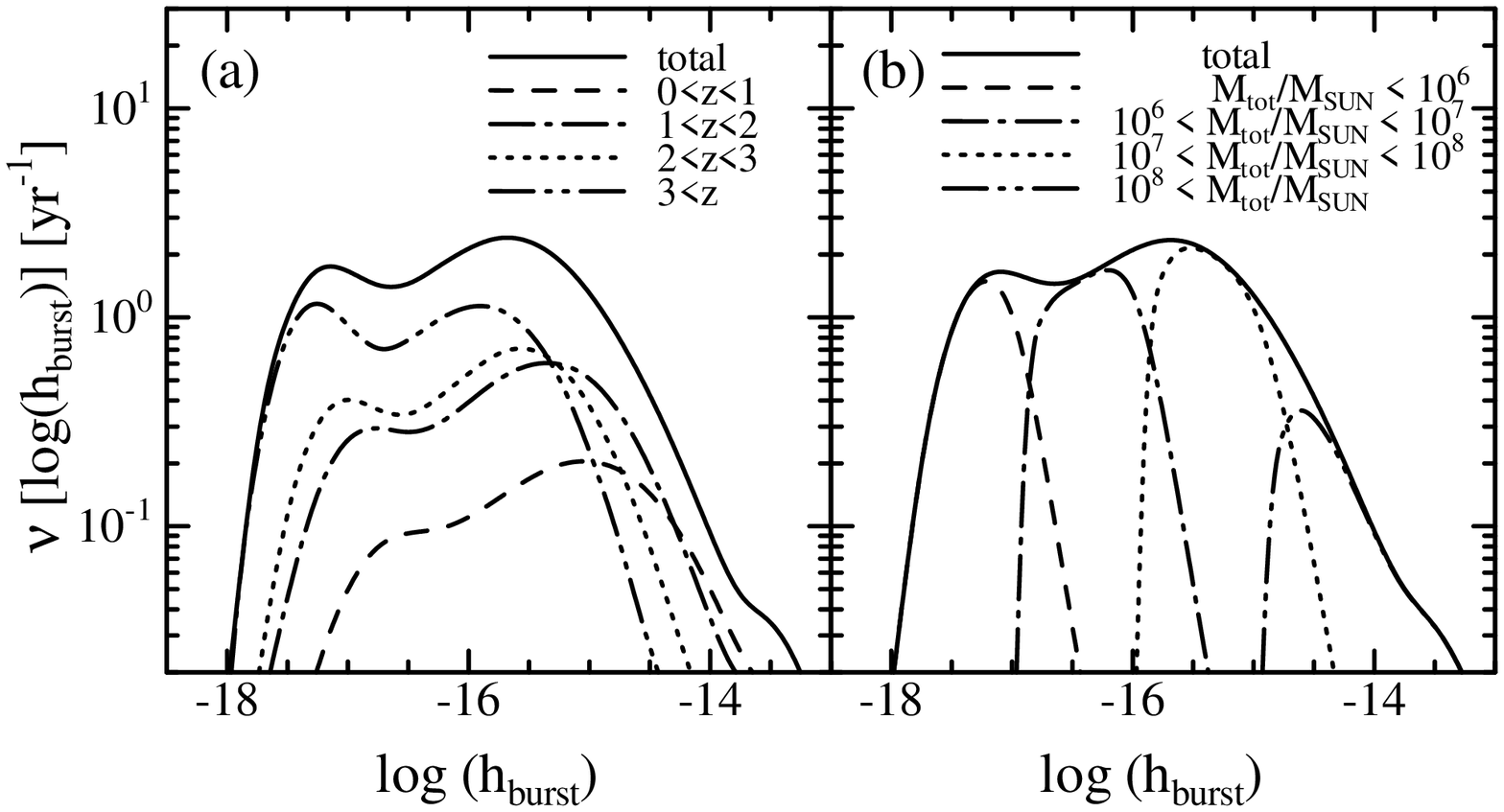}
\caption{Integrated event rate of gravitational wave bursts from SMBH coalescence per
 observers' time  unit a year, $\nu_{\rm burst} [\log(h_{\rm burst})]$,
 in different redshifts (a) and  in different total mass ranges (b).\label{fig4}}
\end{figure*}

\begin{figure*}
\includegraphics[width=95mm]{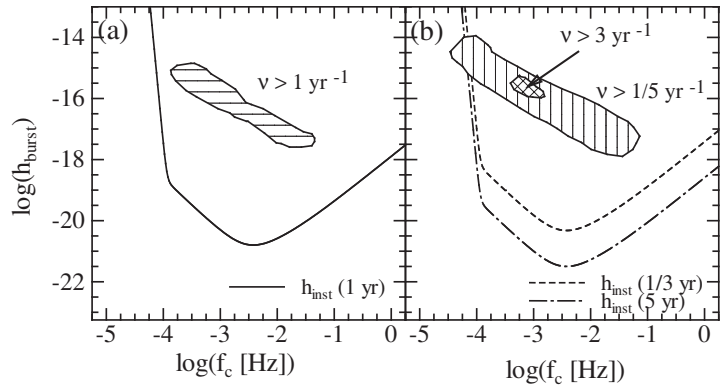}
\caption{Expected signals of gravitational wave bursts from SMBH coalescence. (a)
The horizontally hatched area shows
 the region, $\nu_{\rm burst} [\log(h_{\rm burst}),\log(f_{\rm c})] > 1 {\rm yr}^{-1}$. The solid curve indicates the instrumental noise threshold for one year of {\sl LISA} observations. (b) The vertically hatched area shows the region, $\nu_{\rm burst} > 1/5 {\rm yr}^{-1}$ and the diagnal cross-hatched area show the region, $\nu_{\rm burst} > 3 {\rm yr}^{-1}$. The dot-dashed and the short dashed lines indicates the instrumental noise threshold for $5$ year  and $1/3$ year of {\sl LISA} observations, respectively. 
 \label{fig5}}
\end{figure*}

After an inspiraling phase, SMBHs plunge into a final death inspiral 
 and merge to form a single black hole. We call a set of a 
plunge and a subsequent 
early non-linear ring-down phase as a {\it burst}.
We estimate 
the expected burst event rate per observers' time
using the amplitude of burst gravitational wave given in  \cite{Thorne76}, 
and the SMBH coalescence rate calculated by our SA model, $n_c(M_1,M_2,z)$. 

The characteristic
amplitude of the burst gravitational wave which we finally obtain is
\begin{eqnarray}
h_{\rm burst} &=& \frac{3^{3/4} \epsilon^{1/2} G M_{\rm tot}}{2^{1/2} \pi c^2 D(z)} \nonumber \\
 &=& 7.8 \times 10^{-16} \left(\frac{\epsilon }{0.1} \right)^{1/2} \left(\frac{M_{\rm tot}}{10^8 \ M_{\odot}}\right) \left[\frac{D(z)}{1 \ {\rm Gpc}} \right]^{-1}, \label{eq:ampburst} 
\end{eqnarray}
where $\epsilon$ is the efficiency of the energy release, $D(z)$ is
the comoving distance to the source from the observer and 
 $f_{\rm c}$ is the observed characteristic frequency.
$f_{\rm c}$ from the gravitational
wave burst occurring at $z$ is
\begin{eqnarray}
f_{\rm c} &=&  \frac{c^3}{3^{3/2} GM_{\rm tot} \ (1+z)} \nonumber \\
 &=& 3.9 \times 10^{-4} \left(\frac{M_{\rm tot}}{10^8 \ M_{\odot}}\right)^{-1}
 (1+z)^{-1} {\rm Hz}. \label{eq:burstfreq}
\end{eqnarray}
Then, we obtain $n_{\rm burst} (h_{\rm burst},f_{\rm c}, z) \;dh_{\rm
burst} \;df_{\rm c} \;dz $ , which is the comoving  number density of
gravitational wave burst events occurring at $z$ in a given redshift
interval $dz$ with amplitude $h_{\rm burst} \sim h_{\rm burst} + dh_{\rm
burst}$ and with
 characteristic frequency  $f_{\rm c} \sim f_{\rm c} + df_{\rm c}$.
Therefore, the expected event rates of gravitational wave bursts per observers' time with amplitude $h_{\rm burst} \sim h_{\rm burst} + dh_{\rm burst}$ and
 characteristic frequency  $f_{\rm c} \sim f_{\rm c} + df_{\rm c}$,
 $\nu_{\rm burst}(h_{\rm burst},f_{\rm c})  \;dh_{\rm burst} \;df_{\rm
 c} $ is given by  
\begin{eqnarray}
\nu_{\rm burst}(h_{\rm burst},f_{\rm c}) &=& \int n_{\rm burst} (h_{\rm burst},f_{\rm c}, z) \frac{dV}{dt_0} dz, \label{eq:burstrate}
\end{eqnarray}
where $t_0$ is the observers' time and 
$dV$ is the comoving volume element at $z$.
The integrated event rates of gravitational wave bursts per
 observers' time with amplitude $h_{\rm burst} \sim h_{\rm burst} + dh_{\rm
 burst}$, $\nu_{\rm burst}(h_{\rm burst})  \;dh_{\rm burst}$, 
are given respectively as follows,
\begin{equation}
\nu_{\rm burst}(h_{\rm burst}) = \int \nu_{\rm burst}(h_{\rm burst},f_{\rm c})
d f_{\rm c}. \label{eq:hburstrate}
\end{equation}

In figure \ref{fig4}, we plot the total integrated event rates of
gravitational wave bursts and integrated event rates in
different redshift intervals (fig. \ref{fig4}(a)) and in different total mass intervals in (fig. \ref{fig4}(b)).
Here we set the efficiency of the energy release
$\epsilon = 0.1$, while the precise value of this 
parameter is unknown. In most typical events, a conversion efficiency will
 probably be a few percent.
From equation (\ref{eq:ampburst}), 
one can see that the change of efficiency results in the 
parallel displacement in the horizontal direction
in figure \ref{fig4}. 
The shape of $\nu_{\rm burst}(h_{\rm burst})$ reflects the black hole
mass functions and the SMBH coalescence rates, which depend on the
complex galaxy formation processes.
In figure \ref{fig4}  (a), one can notice that 
there are two peaks in the event rate in terms of $h_{\rm burst}$.  
A peak at $h_{\rm burst} \sim 10^{-17}$ corresponds to bursts 
from SMBH binaries with $M_{\rm
tot} < 10^6 M_{\odot}$ whose total number
is the largest at high redshift $z>3$.
Another peak at $h_{\rm burst} \sim 10^{-15}$
corresponds to bursts from SMBH 
binaries with  $ 10^7M_{\odot} < M_{\rm tot} < 10^8 M_{\odot}$ 
whose coalescence probability is the largest at low redshift $z<3$.
Figure \ref{fig4} (b) indicates that burst signals with large amplitude 
($h_{\rm burst} \gtrsim 10^{-15}$) correspond to coalescence of
``massive'' SMBH binaries   
with $M_{\rm tot} \gtrsim 10^8 M_{\odot}$ occurring at $z
\lesssim 1$. This is because the distance from the earth 
to SMBHs at low redshift is
shorter and the mass of SMBHs at low redshift is larger. 
On the other hand,  burst signals with small  amplitude 
($h_{\rm burst} \lesssim 10^{-17}$) corresponds to
coalescence events of 
``less massive''  SMBH
binaries with $M_{\rm tot} \lesssim 10^7 M_{\odot}$ 
occurring at $z \gtrsim 2$. These events dominate the 
expected burst event rate provided that the sensitivity 
of the detector is sufficiently good.
This feature is quite important because it breaks the degeneracy
between mass and distance.

Figure \ref{fig5} shows that the expected region for signal of
gravitational wave bursts and 
the instrumental noise threshold for {\sl LISA}, $h_{\rm
inst}$.
We  compute $h_{\rm inst}$
from the fitting formula for the spectral instrumental noise density of {\sl LISA} \citep{Hughes02}. The expected region for $\nu_{\rm burst} [\log(h_{\rm
burst}),\log(f_{\rm c})] > 1 \ {\rm yr}^{-1}$ is above this instrumental
noise threshold. For comparison, we show the region for  $\nu_{\rm burst} [\log(h_{\rm
burst}),\log(f_{\rm c})] > 1/5 \ {\rm yr}^{-1}$ and  $\nu_{\rm burst} [\log(h_{\rm
burst}),\log(f_{\rm c})] > 3 \ {\rm yr}^{-1}$
 in figure \ref{fig5} (b).

From figure \ref{fig4} and \ref{fig5}, we conclude that the
{\sl LISA} can 
detect intense bursts of gravitational waves at a rate of  
 $\sim 1.0  {\rm yr}^{-1}$ assuming
that dominant part of these burst events occur at
$z \gtrsim 2$. 
Even in the case of $\epsilon = 0.001$, the
{\sl LISA} can detect intense bursts of gravitational
waves in one year observation, since 
$h_{\rm burst} \propto \epsilon^{1/2} $. 
In addition, we find that large
amplitude $h_{\rm burst} \sim 10^{-15}$
signals correspond to coalescence events of massive 
SMBH binaries $M_{\rm tot} \sim 10^8
M_{\odot}$
at low redshift $ z \lesssim 1$ and small amplitude $h_{\rm
 burst} \sim 10^{-17}$  signals correspond to less massive 
SMBH binaries $M_{\rm tot} \sim 10^6
M_{\odot}$ at high redshift $ z \gtrsim 3$.

\section{Summary \& Conclusions}
In this study, we have estimated the coalescence rate of SMBH binaries in the
centers of galaxies using a new SA model of
galaxy and quasar formation \cite{Enoki03}. Then, 
we calculated the spectrum of the gravitational wave
background from inspiraling SMBH binaries and estimated the expected amplitudes and 
event rates of intense bursts of
gravitational waves from coalescing SMBH binaries \cite{Enoki04}.

Our SA model includes dynamical friction and random collision as galaxy merging mechanisms, and assumes that a SMBH is fueled by accretion of cold gas
during a  major merger of galaxies leading to a spheroid formation,
and that SMBHs coalesce simultaneously when host galaxies
merge. Many previous other studies have paid attention to only SMBH growth
and did not take galaxy formation processes into account. For
investigating the relations between SMBH growth and galaxy formation
processes, SA methods of galaxy and SMBH formation are suitable tools 
 \citep[e.g.][]{KH00,Enoki03}.   

We have found that the gravitational wave background radiation spectrum for
$f \lesssim 1 \mu {\rm Hz}$ has a characteristic
strain $h_c(f) \sim 10^{-16} (f/1 \mu {\rm Hz})^{-2/3}$ just below the
 detection limit from the current measurements of the pulsar timing.
The slope of the spectrum for $f \gtrsim 1 \mu {\rm Hz}$ gets steep
owing to the upper limit in frequency set by 
the radius of the ISCO. 
The stochastic 
background radiation mainly comes from
inspiraling SMBH binaries at $0<z<1$. Therefore, the background radiation
can probe inspiraling SMBH binaries at low redshift.

We have also found that {\sl LISA} might detect
 intense bursts of gravitational waves owing to the SMBH coalescence events
at a rate $0.1 \sim 1.0 {\rm ~yr}^{-1}$ and that the main
contribution to the event rate comes from 
SMBH binary coalescence events at high redshift $z \gtrsim 2$.
Our model predicts that burst signals with a large amplitude 
correspond to coalescence of large mass SMBH  binaries 
at low redshift while those with a small amplitude 
correspond to coalescence of small mass SMBH binaries at high redshift. 
This prediction can be tested by future measurements 
of the amplitude and the phase evolution in 
gravitational waves from inspiraling SMBH binaries \citep{Hughes02}. 
Comparing these
predictions with observations in future, 
we can put a stringent 
constraint on SMBH formation and evolution models.

\bigskip 
\begin{acknowledgments}
 We thank K.S. Thorne and S. Hughes for 
useful information.
MN acknowledges Research Fellowships of the Japan Society
for the Promotion of Science for Young Scientists (No.00207).
NS is supported by Japanese Grant-in-Aid for Science Research Fund of
the Ministry of Education, No.14340290.
\end{acknowledgments}

\bigskip 

\end{document}